\documentclass[aps,prl,twocolumn,showpacs,groupedaddress,floatfix]{revtex4}  
\usepackage{times}
\usepackage{graphicx}  
\usepackage{dcolumn}   
\usepackage{bm}        
\usepackage{amssymb}   
\usepackage{xspace}

\begin{document}
\newcommand{\dzero} {D\O\xspace} \newcommand{\ttbar}
{\ensuremath{t\bar{t}}\xspace} \newcommand{\ppbar}
{\ensuremath{p\bar{p}}\xspace} \newcommand{\thetad}
{\ensuremath{\theta^{*}}\xspace}
\newcommand{\costheta}{\ensuremath{\cos\thetad}\xspace}
\newcommand{\ljets}{\ensuremath{\ell +}jets\xspace}
\newcommand{\mujets}{\ensuremath{\mu+jets}\xspace}
\newcommand{\ejets}{\ensuremath{e+jets}\xspace}
\newcommand{\fplus}{\ensuremath{f_{+}}\xspace}
\newcommand{\fminus}{\ensuremath{f_{-}}\xspace}
\newcommand{\fzero}{\ensuremath{f_{0}}\xspace}
\newcommand{\wjets}{\ensuremath{W+}jets\xspace} 
\newcommand{\met}{\mbox{$\not\!\!E_T$}\xspace} 
\newcommand{\tld}{\ensuremath{{\cal D}}\xspace} 
\newcommand{\tWb}{\ensuremath{t \rightarrow Wb}\xspace} 
\newcommand{\SM}{standard model\xspace} 
\newcommand{\MC}{Monte Carlo\xspace} \newcommand{\CL}{C.L.\xspace}
\newcommand{\deteta}{\ensuremath{\eta_{\mathrm{det}}}}
\newcommand{\alpgen}{{\sc{alpgen}}\xspace}
\newcommand{\pythia}{{\sc{pythia}}\xspace}
\hyphenation{an-aly-sis}




\hspace{5.2in} \mbox{FERMILAB-PUB-06-345-E}
\vspace*{1.5cm}

\title{Measurement of the $W$ boson helicity in top quark decay at D\O}


%
\author{                                                                      
V.M.~Abazov,$^{35}$                                                           
B.~Abbott,$^{75}$                                                             
M.~Abolins,$^{65}$                                                            
B.S.~Acharya,$^{28}$                                                          
M.~Adams,$^{51}$                                                              
T.~Adams,$^{49}$                                                              
M.~Agelou,$^{17}$                                                             
E.~Aguilo,$^{5}$                                                              
S.H.~Ahn,$^{30}$                                                              
M.~Ahsan,$^{59}$
% ...
\\                                                                            
\vskip 0.30cm                                                                 
\centerline{(D\O\ Collaboration)}                                             
\vskip 0.30cm                                                                 
}                                                                             
\affiliation{                                                                 
\centerline{$^{1}$Universidad de Buenos Aires, Buenos Aires, Argentina}       
\centerline{$^{2}$LAFEX, Centro Brasileiro de Pesquisas F{\'\i}sicas,         
                  Rio de Janeiro, Brazil}                                     
\centerline{$^{3}$Universidade do Estado do Rio de Janeiro,                   
                  Rio de Janeiro, Brazil}                                     
\centerline{$^{4}$Instituto de F\'{\i}sica Te\'orica, Universidade            
                  Estadual Paulista, S\~ao Paulo, Brazil}                     
\centerline{$^{5}$University of Alberta, Edmonton, Alberta, Canada,           
                  Simon Fraser University, Burnaby, British Columbia, Canada,}
\centerline{York University, Toronto, Ontario, Canada, and                    
                  McGill University, Montreal, Quebec, Canada}                
\centerline{$^{6}$University of Science and Technology of China, Hefei,       
                  People's Republic of China}                                 
\centerline{$^{7}$Universidad de los Andes, Bogot\'{a}, Colombia}             
\centerline{$^{8}$Center for Particle Physics, Charles University,            
                  Prague, Czech Republic}                                     
\centerline{$^{9}$Czech Technical University, Prague, Czech Republic}         
\centerline{$^{10}$Center for Particle Physics, Institute of Physics,         
                   Academy of Sciences of the Czech Republic,                 
                   Prague, Czech Republic}                                    
\centerline{$^{11}$Universidad San Francisco de Quito, Quito, Ecuador}        
\centerline{$^{12}$Laboratoire de Physique Corpusculaire, IN2P3-CNRS,         
                   Universit\'e Blaise Pascal, Clermont-Ferrand, France}      
\centerline{$^{13}$Laboratoire de Physique Subatomique et de Cosmologie,      
                   IN2P3-CNRS, Universite de Grenoble 1, Grenoble, France}    
\centerline{$^{14}$CPPM, IN2P3-CNRS, Universit\'e de la M\'editerran\'ee,     
                   Marseille, France}                                         
\centerline{$^{15}$IN2P3-CNRS, Laboratoire de l'Acc\'el\'erateur              
                   Lin\'eaire, Orsay, France}                                 
\centerline{$^{16}$LPNHE, IN2P3-CNRS, Universit\'es Paris VI and VII,         
                   Paris, France}                                             
\centerline{$^{17}$DAPNIA/Service de Physique des Particules, CEA, Saclay,    
                   France}                                                    
\centerline{$^{18}$IPHC, IN2P3-CNRS, Universit\'e Louis Pasteur, Strasbourg,  
                    France, and Universit\'e de Haute Alsace,                 
                    Mulhouse, France}                                         
\centerline{$^{19}$Institut de Physique Nucl\'eaire de Lyon, IN2P3-CNRS,      
                   Universit\'e Claude Bernard, Villeurbanne, France}         
\centerline{$^{20}$III. Physikalisches Institut A, RWTH Aachen,               
                   Aachen, Germany}                                           
\centerline{$^{21}$Physikalisches Institut, Universit{\"a}t Bonn,             
                   Bonn, Germany}                                             
\centerline{$^{22}$Physikalisches Institut, Universit{\"a}t Freiburg,         
                   Freiburg, Germany}                                         
\centerline{$^{23}$Institut f{\"u}r Physik, Universit{\"a}t Mainz,            
                   Mainz, Germany}                                            
\centerline{$^{24}$Ludwig-Maximilians-Universit{\"a}t M{\"u}nchen,            
                   M{\"u}nchen, Germany}                                      
\centerline{$^{25}$Fachbereich Physik, University of Wuppertal,               
                   Wuppertal, Germany}                                        
\centerline{$^{26}$Panjab University, Chandigarh, India}                      
\centerline{$^{27}$Delhi University, Delhi, India}                            
\centerline{$^{28}$Tata Institute of Fundamental Research, Mumbai, India}     
\centerline{$^{29}$University College Dublin, Dublin, Ireland}                
\centerline{$^{30}$Korea Detector Laboratory, Korea University,               
                   Seoul, Korea}                                              
\centerline{$^{31}$SungKyunKwan University, Suwon, Korea}                     
\centerline{$^{32}$CINVESTAV, Mexico City, Mexico}                            
\centerline{$^{33}$FOM-Institute NIKHEF and University of                     
                   Amsterdam/NIKHEF, Amsterdam, The Netherlands}              
\centerline{$^{34}$Radboud University Nijmegen/NIKHEF, Nijmegen, The          
                  Netherlands}                                                
\centerline{$^{35}$Joint Institute for Nuclear Research, Dubna, Russia}       
\centerline{$^{36}$Institute for Theoretical and Experimental Physics,        
                   Moscow, Russia}                                            
\centerline{$^{37}$Moscow State University, Moscow, Russia}                   
\centerline{$^{38}$Institute for High Energy Physics, Protvino, Russia}       
\centerline{$^{39}$Petersburg Nuclear Physics Institute,                      
                   St. Petersburg, Russia}                                    
\centerline{$^{40}$Lund University, Lund, Sweden, Royal Institute of          
                   Technology and Stockholm University, Stockholm,            
                   Sweden, and}                                               
\centerline{Uppsala University, Uppsala, Sweden}                              
\centerline{$^{41}$Physik Institut der Universit{\"a}t Z{\"u}rich,            
                   Z{\"u}rich, Switzerland}                                   
\centerline{$^{42}$Lancaster University, Lancaster, United Kingdom}           
\centerline{$^{43}$Imperial College, London, United Kingdom}                  
\centerline{$^{44}$University of Manchester, Manchester, United Kingdom}      
\centerline{$^{45}$University of Arizona, Tucson, Arizona 85721, USA}         
\centerline{$^{46}$Lawrence Berkeley National Laboratory and University of    
                   California, Berkeley, California 94720, USA}               
\centerline{$^{47}$California State University, Fresno, California 93740, USA}
\centerline{$^{48}$University of California, Riverside, California 92521, USA}
\centerline{$^{49}$Florida State University, Tallahassee, Florida 32306, USA} 
\centerline{$^{50}$Fermi National Accelerator Laboratory,                     
            Batavia, Illinois 60510, USA}                                     
\centerline{$^{51}$University of Illinois at Chicago,                         
            Chicago, Illinois 60607, USA}                                     
\centerline{$^{52}$Northern Illinois University, DeKalb, Illinois 60115, USA} 
\centerline{$^{53}$Northwestern University, Evanston, Illinois 60208, USA}    
\centerline{$^{54}$Indiana University, Bloomington, Indiana 47405, USA}       
\centerline{$^{55}$University of Notre Dame, Notre Dame, Indiana 46556, USA}  
\centerline{$^{56}$Purdue University Calumet, Hammond, Indiana 46323, USA}    
\centerline{$^{57}$Iowa State University, Ames, Iowa 50011, USA}              
\centerline{$^{58}$University of Kansas, Lawrence, Kansas 66045, USA}         
\centerline{$^{59}$Kansas State University, Manhattan, Kansas 66506, USA}     
\centerline{$^{60}$Louisiana Tech University, Ruston, Louisiana 71272, USA}   
\centerline{$^{61}$University of Maryland, College Park, Maryland 20742, USA} 
\centerline{$^{62}$Boston University, Boston, Massachusetts 02215, USA}       
\centerline{$^{63}$Northeastern University, Boston, Massachusetts 02115, USA} 
\centerline{$^{64}$University of Michigan, Ann Arbor, Michigan 48109, USA}    
\centerline{$^{65}$Michigan State University,                                 
            East Lansing, Michigan 48824, USA}                                
\centerline{$^{66}$University of Mississippi,                                 
            University, Mississippi 38677, USA}                               
\centerline{$^{67}$University of Nebraska, Lincoln, Nebraska 68588, USA}      
\centerline{$^{68}$Princeton University, Princeton, New Jersey 08544, USA}    
\centerline{$^{69}$State University of New York, Buffalo, New York 14260, USA}
\centerline{$^{70}$Columbia University, New York, New York 10027, USA}        
\centerline{$^{71}$University of Rochester, Rochester, New York 14627, USA}   
\centerline{$^{72}$State University of New York,                              
            Stony Brook, New York 11794, USA}                                 
\centerline{$^{73}$Brookhaven National Laboratory, Upton, New York 11973, USA}
\centerline{$^{74}$Langston University, Langston, Oklahoma 73050, USA}        
\centerline{$^{75}$University of Oklahoma, Norman, Oklahoma 73019, USA}       
\centerline{$^{76}$Oklahoma State University, Stillwater, Oklahoma 74078, USA}
\centerline{$^{77}$Brown University, Providence, Rhode Island 02912, USA}     
\centerline{$^{78}$University of Texas, Arlington, Texas 76019, USA}          
\centerline{$^{79}$Southern Methodist University, Dallas, Texas 75275, USA}   
\centerline{$^{80}$Rice University, Houston, Texas 77005, USA}                
\centerline{$^{81}$University of Virginia, Charlottesville,                   
            Virginia 22901, USA}                                              
\centerline{$^{82}$University of Washington, Seattle, Washington 98195, USA}  
}                                                                             

%


\date{September 25, 2006}

\begin{abstract}


  We present a measurement of the fraction $f_+$ of right-handed $W$
  bosons produced in top quark decays, based on a candidate sample of
  $t\bar{t}$ events in the $\ell +$jets and dilepton decay channels 
  corresponding to an
  integrated luminosity of $370$~pb$^{-1}$ collected by the D\O\
  detector at the Fermilab Tevatron $p\bar{p}$ Collider at
  $\sqrt{s}=1.96$~TeV.  We reconstruct the
  decay angle $\theta^*$ for each lepton.  By
  comparing the $\cos\theta^*$ distribution from the data with those for
  the expected background and signal for various values of $f_+$, we
  find $f_+=0.056\pm0.080\thinspace\mathrm{ (stat) }\pm0.057\thinspace\mathrm{ (syst) }$.
  ($f_+ < 0.23$ at 95\% C.L.),  consistent with the standard model prediction 
  of $f_+=3.6\times10^{-4}$.

\end{abstract}


\maketitle 



The top quark is by far the heaviest of the known fermions and is
the only one that has a Yukawa coupling of order unity to the Higgs boson
in the \SM.   
We search for evidence of new physics
in \tWb decay
by measuring the helicity of the $W$ boson.
In the \SM, the top quark decays via the $V-A$ charged current interaction,
almost always 
to a $W$ boson and a $b$ quark. 
For any linear combination of $V$ and $A$ currents at the \tWb vertex, the 
fraction \fzero of longitudinally-polarized $W$ bosons is 
$0.697\pm0.012$~\cite{fzero} at the world average top quark mass $m_t$ of 
$172.5 \pm 2.3~{\rm GeV}$~\cite{WAtopmass}.


In this analysis, we fix \fzero at $0.70$ and measure the
positive helicity 
fraction \fplus.
  In the \SM, 
\fplus 
is predicted
at next-to-leading order
to be $3.6\times10^{-4}$ ~\cite{fischer1}.
  A measurement of \fplus that differs significantly from
this value would be an unambiguous indication of new physics.
For example, an \fplus value of $0.30$ would 
indicate a purely $V+A$ charged current interaction.  

  Measurements of the $b \rightarrow s\gamma$ decay rate have indirectly limited
the $V+A$ contribution in top quark decays to less than a few percent~\cite{sbg1}.
Direct measurements of the $V+A$ contribution are still necessary because
the limit from $b \rightarrow s\gamma$  assumes that the electroweak 
penguin contribution is dominant.  Direct 
measurements of the longitudinal fraction
found $\fzero=0.91\pm0.39$~\cite{helicityCDF2000} and $\fzero=0.56\pm0.31$~\cite{helicityD0}. 
Direct measurements of \fplus have set limits of 
$\fplus<0.18$~\cite{helicityCDF2004}, $\fplus<0.24$~\cite{helicityCDF2005},
and $\fplus<0.25$~\cite{helicityD02005} at the $95\%$ \CL  The analysis
presented in this Letter improves upon that reported in Ref.~\cite{helicityD02005}
by using a larger data set, including the dilepton decay channel of the
$t\bar{t}$ pair, and 
employing enhanced analysis techniques.

The angular distribution of the down-type decay products of the $W$ boson
(charged lepton or $d$, $s$ quark)
in the rest frame of the $W$ boson can be described by 
introducing the decay angle \thetad of the down-type particle 
with respect to the top quark direction.
The dependence of the distribution of
$\cos\thetad$ on \fplus,
\begin{equation}
\omega(c_{\thetad}) \propto 2(1-c^2_{\thetad})\fzero+
(1-c_{\thetad})^2\fminus\
                +(1+c_{\thetad})^2\fplus\,
\end{equation}
where $c_\thetad = \cos\thetad$, forms the basis for our measurement.
We proceed by selecting a data sample
enriched in \ttbar events, reconstructing the four vectors of the two
top quarks and their decay products, and then
calculating \costheta. 
This distribution in \costheta is compared with 
templates for different \fplus values, suitably corrected for background and 
reconstruction
effects, using a binned maximum
likelihood method.  In the $\ell +$jets channel, the kinematic 
reconstruction is done with a fit that constrains the $W$ boson
mass to its measured value and the top quark mass to 175 GeV, while in 
the dilepton channel, the kinematics are solved algebraically with the top
quark mass fixed to 172.5 GeV.

The D\O\ detector~\cite{d0det} comprises three main systems: the central tracking system,
the calorimeters, and the muon system. 
The central-tracking system 
is located within a 2~T
solenoidal magnet. 
The next layer of 
detection involves three liquid-argon/uranium calorimeters: a central 
section covering pseudorapidities~\cite{def_eta} $|\eta|\lesssim 1$, and two end
calorimeters extending coverage to $|\eta|\approx 4$, all housed 
in separate cryostats. 
The muon system is located outside the calorimetry, and consists of a 
layer of tracking detectors and scintillation trigger counters 
before 1.8~T toroids, followed by two similar layers after
the toroids.  


This measurement uses a data sample recorded with the \dzero
experiment and corresponds to an integrated luminosity of about 
$370$~pb$^{-1}$
of \ppbar collisions at $\sqrt{s}=1.96$~TeV.
The data sample consists of \ttbar candidate
events from the \ljets decay channel 
$t\bar{t}\rightarrow W^+W^-b\bar{b}\rightarrow \ell\nu qq^{\prime}b\bar{b}$
 and the dilepton channel
$t\bar{t}\rightarrow W^+W^-b\bar{b}\rightarrow \ell\nu\ell^{\prime}\nu^{\prime}b\bar{b}$, where $\ell$ and $\ell^{\prime}$ are electrons or muons.  The \ljets final state is
characterized by one charged lepton, at least four jets (two of
which are $b$ jets), and significant missing transverse energy (\met).  The
dilepton final state is characterized by two charged leptons of opposite sign,
at least two jets, and significant \met.

We simulate \ttbar signal events with $m_t = 172.5$ GeV  
for different
values of $f_+$
with the \alpgen Monte Carlo (MC) program~\cite{alpgen} for the 
parton-level process (leading order) and \pythia~\cite{pythia} for 
gluon radiation and subsequent
hadronization. 
As the
interference term between $V-A$ and $V+A$ is suppressed by the small
mass of the $b$ quark and is therefore
negligible~\cite{interference}, samples with $f_+=0.00$ and 
$f_+=0.30$ are used to
create \costheta templates for any $f_+$ value by a linear interpolation of the
templates.  

The MC samples used to model background events with real leptons
are also generated
using \alpgen and \pythia.
Backgrounds in the \ljets channel arise predominantly from \wjets production 
and multijet
production where one of the jets is misidentified as a lepton and
spurious \met appears due to mismeasurement of the transverse energy
in the event.  

The \ljets event selection~\cite{xsec_topo} requires an isolated lepton 
($e$ or $\mu$) with
transverse momentum 
$p_T>20$~$\mathrm{GeV}$, no other lepton with $p_T>15$~$\mathrm{GeV}$
in the event, $\met>20$~GeV, and at least four jets.
Electrons are
required to have $|\eta|<1.1$ and are identified by
their energy deposition and isolation in the calorimeter, their transverse and
longitudinal shower shapes, and
information from the tracking system. 
Also, a discriminant
combining the above information must be consistent with the expectation for a
high-$p_T$ isolated electron~\cite{xsec_topo}.
Muons are identified using information from the muon and tracking 
systems, and must satisfy isolation requirements based on the energies of calorimeter clusters and
the momenta of tracks around the muon. They are required to have $|\eta|<2.0$ and to be
isolated from jets. 
Jets are reconstructed using the Run II mid-point cone algorithm with cone 
radius 0.5~\cite{jetreco}, and are required to have rapidity  $|y|<2.5$ and $p_T>20$~$\mathrm{GeV}$.

We determine the number of multijet
background events $N_{mj}$ from the data, using the technique described
in Ref.~\cite{xsec_topo}.
We calculate $N_{mj}$ for each bin in the
\costheta distribution from the data sample to obtain the
multijet \costheta templates.


To discriminate between \ttbar pair production and background, a
discriminant $\tld$ with values in the range 0 to 1 is calculated using input 
variables which exploit
differences in kinematics and jet flavor.  The kinematic variables 
considered are:
$H_T$ (defined as the scalar sum of the
jet $p_T$ values), the minimum dijet mass of the jet pairs $m_{jj{\rm min}}$,
 the
$\chi^2$ from the kinematic fit, 
the difference in azimuthal angle $\Delta \phi$ between the lepton and \met 
directions, 
and aplanarity ${\cal A}$ and sphericity ${\cal S}$~\cite{apla} (calculated 
from the four
leading jets and the lepton). 
Only the four leading jets in $p_T$ are considered in computing these
variables.

We utilize the fact that background jets arise mostly from light quarks or
gluons while two of the jets in \ttbar events arise from $b$ quarks by considering
the impact parameters with respect to the primary vertex of all tracks within the jet cone.  Based on these values, we calculate the 
probability $P_{PV}$ 
for each
jet to originate from the primary vertex.  We then average the two smallest 
$P_{PV}$ values to form a continuous variable $\langle P_{PV} \rangle$ that 
tends to be small for
\ttbar events and large for backgrounds.  This approach results in similar 
background discrimination but better efficiency than applying a simple 
cut on $P_{PV}$.

The discriminant is built separately for the $e+$jets and
$\mu+$jets channels, using the method described in Refs.~\cite{xsec_topo,run1_topmass}.  Background events tend to have \tld values near 0, while $t\bar{t}$
events tend to have values near 1.
%
%
%
We consider all possible combinations of the above variables for use in the 
discriminant, and all possible requirements on the \tld value,
 and choose the variables and \tld criterion that give the smallest expected
uncertainty on \fplus.  In the $e+$jets channel, ${\cal S}$, $H_T$, 
 $\langle P_{PV} \rangle$, and $\chi^2$ are used, and 
\tld is required to be $>0.65$.  In the $\mu+$jets channel, 
${\cal A}$, $H_T$, $m_{jj {\rm min}}$, $\langle P_{PV} \rangle$,  
$\chi^2$, and $\Delta \phi$ are used, and \tld is required to be $>0.80$.   
In both channels the efficiency for $t\bar{t}$ events to satistfy the
\tld requirement is independent of the value of $f_+$.


We then perform a binned Poisson maximum likelihood fit to compare
the observed distribution of events in \tld  
 to the sum of the
distributions expected from \ttbar, \wjets, and multijet events.
$N_{mj}$ 
is constrained to the expected value within the known uncertainty.
The likelihood is then
maximized with respect to the numbers of \ttbar, \wjets, and
multijet events,  
which are multiplied by the appropriate efficiency for the \tld selection 
to determine the composition
of the sample used for measuring \costheta.

In the dilepton channel, backgrounds arise from processes such as $WW+$jets 
or $Z+$jets.  These
processes are either rare or require false \met from mismeasurement
of jet and lepton energy, allowing a good signal to background ratio to be 
attained using 
only kinematic selection criteria.  The selection is detailed 
in Ref.~\cite{ref:ll_xsec}.  Events are required to have two leptons with opposite
charge and $p_T > 15$ GeV and two or more jets with $p_T > 20 $ GeV and
$|y| < 2.5$.  Additional criteria are applied in the $ee$ and $\mu\mu$ channels
to suppress $Z\rightarrow\ell\ell$, and in the $e\mu$ channel the sum of the two
leading jet $p_T$'s and the leading lepton $p_T$ must be greater than 122 GeV.  
We place a more stringent requirement on electron identification than
is used in Ref.~\cite{ref:ll_xsec}. 

Table~\ref{tab_sig_bkg_numbers} lists the composition of
each sample as well as the number of observed events in the data. 
We observe a disparity between the number of \ttbar\ events in 
the $e+$jets channel and $\mu$+jets channel, which is unexpected
since the selection efficiencies for the two channels are similar.  The 
statistical significance
of the discrepancy in the event distribution is slightly above 
$2\sigma$.  The disparity appears to be
a feature of the data sample used in this analysis, as it occurs regardless
of the choice of variables used to define ${\cal D}$.  Further, it has no
direct impact on this analysis, which relies only upon the distribution of 
events in \costheta.  

 \begin{table}
\caption{\label{tab_sig_bkg_numbers}%
Number of events observed in each \ttbar decay channel, the
background level as determined by a fit to the \tld distribution in
the \ljets channels and the expectation from the background production
rate and selection efficiency in the dilepton channels, and the expected
signal yield assuming standard model $t\bar{t}$ production with a top quark 
mass of 175 GeV. 
}
\begin{tabular}{lccc}
\hline \hline
       & Observed & Background & Expected $t\bar{t}$ \\
\hline
$e$ + jets    &   51 & $5.3 \pm 0.9$ & 32.9 \\
$\mu$ + jets  &   19 &  $3.3 \pm 0.4$  & 26.4                \\
$e\mu$        &    15            &    $2.2 \pm 0.6$  & 8.9            \\
$ee$          &     4           &     $0.8 \pm 0.2$  & 3.3          \\
$\mu\mu$      &     1           &     $0.4 \pm 0.1$  & 2.4          \\ \hline \hline

\end{tabular}
\end{table}

\begin{figure}
\includegraphics[trim=0 15 0 7,scale=0.75]{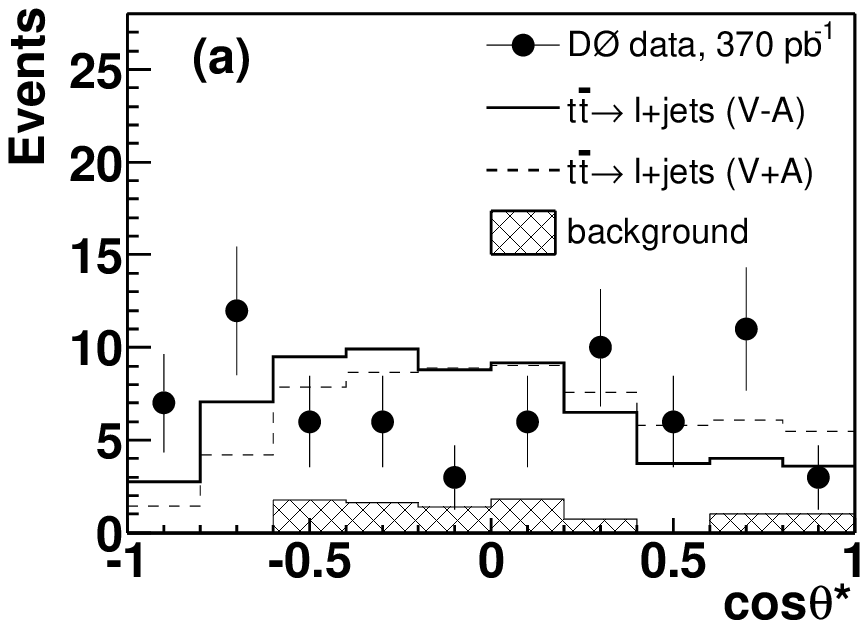} \\ \vskip 0.35cm 
\includegraphics[trim=0 15 0 7,scale=0.75]{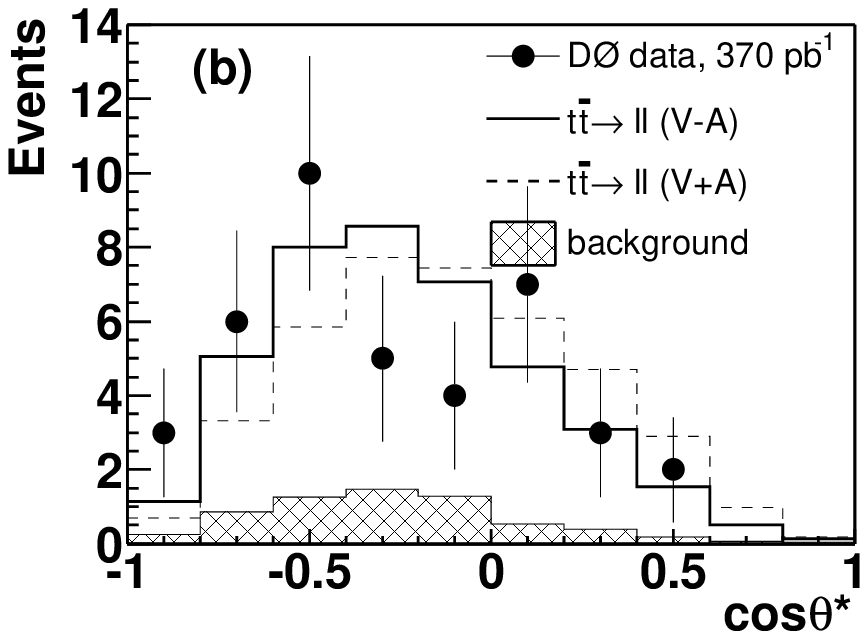}
\caption{\label{fig_cost_ljets}\costheta distribution observed in (a) \ljets 
and (b) dilepton events. The \SM prediction 
is shown as the solid line, while a model with a pure $V+A$ interaction would result in the distribution given by the dashed line.}
\end{figure}

The top quark and $W$ boson four-momenta in the selected \ljets events
are reconstructed using a
kinematic fit which is subject to these constraints: two jets
must form the invariant mass of the $W$ boson, the lepton and the \met
together with the neutrino $p_z$ component must form the invariant
mass of the $W$ boson, and the masses of the two reconstructed top quarks
must be $175$~$\mathrm{GeV}$.
Among the twelve possible jet combinations, the solution with the
minimal $\chi^2$ from the kinematic fit is chosen; MC studies show
this yields the correct solution in about $60\%$ of all cases.
The
\costheta distribution obtained in the \ljets data after the full selection and
compared to standard and $V+A$ model expectations is
shown in Fig.~\ref{fig_cost_ljets}(a).

Dilepton events are rarer than \ljets events, but have the advantage that
\costheta can be calculated for each lepton, thus providing two
measurements per event. 
The presence of two neutrinos in the dilepton final state makes the system
kinematically underconstrained.  However, if a top quark mass is assumed, the 
kinematics
can be solved algebraically with a four-fold ambiguity in addition to the 
two-fold ambiguity in pairing jets with leptons.  For each lepton, we 
calculate the
value of \costheta resulting from each solution with each of the two leading 
jets associated with the lepton.  To account for detector resolution we
repeat the above procedure 100 times, fluctuating the jet and lepton energies 
within their resolutions for each iteration.
The average of these values is taken as 
the \costheta for that lepton.  The
\costheta distribution obtained in dilepton data is
shown in Fig.~\ref{fig_cost_ljets}(b).

%
We  compute the binned Poisson likelihood $L(\fplus)$ for the data to be consistent
with the sum of signal and background templates at
each of seven chosen \fplus values.  
A parabola is fit to the
$-\ln[L(\fplus)]$ points to determine the likelihood as a function
of $\fplus$. 

%
%



Systematic uncertainties are evaluated in ensemble tests
by varying the parameters (see Table~\ref{tab_systematic})
which can affect the shapes of the \costheta distributions
or the relative contribution from signal and background sources.  
Ensembles are formed by drawing events from a model with the parameter 
under study varied.  These are compared to the standard \costheta templates in a maximum 
likelihood fit.  The average shift in the resulting \fplus value is taken 
as the systematic uncertainty and is shown in Table~\ref{tab_systematic}.
The total systematic uncertainty is then
taken into account in the likelihood by convoluting the latter with a 
Gaussian with a width that corresponds
to the total systematic uncertainty. The dominant uncertainties arise from 
the uncertainties on
the top quark mass and 
on the jet energy scale (JES). The mass of the top quark is
 varied by $\pm2.3$~$\mathrm{GeV}$
and the JES by $\pm1\sigma$ around their nominal values.

The statistical uncertainty on the \costheta templates is taken as a 
systematic uncertainty estimated by fluctuating the templates  
according to their 
statistical uncertainty, and noting the RMS of the resulting distribution 
 when fitting to the data. 

The effect of gluon radiation in the modeling of \ttbar
events is studied with an alternate MC sample that
includes $t\bar{t}$ events generated with an additional hard parton
by \alpgen.  These events are mixed with the standard $t\bar{t}$  events
according to the ratio of the leading order cross sections for these
two processes.
Effects of the chosen factorization scale $Q$ in the generation of the $W$+jets
events are evaluated using a sample generated with a different choice 
of $Q$.
The systematic uncertainty on the jet flavor composition in the 
$W$+jets background is derived using alternate MC samples in which the fraction
of $b$ and $c$ jets are varied by 20\% about the nominal value \cite{ref:ljet_btag_xsec}.
The difference found between the input 
\fplus value and the
reconstructed \fplus value in ensemble tests is taken as the systematic 
uncertainty on the calibration of the analysis.
\begin{table}
\caption{\label{tab_systematic}Systematic uncertainties on \fplus for the 
two channels and for their combination.}
\begin{tabular}{lccc}
\hline \hline
Source & \ljets & Dilepton & Combined\\
\hline
Jet energy scale    & 0.038 & 0.039 & 0.038\\
Top quark mass      & 0.019 & 0.028 & 0.021\\
Template statistics & 0.037 & 0.024 & 0.028\\ 
\ttbar model        & 0.006 & 0.018 & 0.009\\
Background model    & 0.007 & 0.007 & 0.005\\
Heavy flavor fraction & 0.018 & --  & 0.015\\
Calibration         & 0.018 & 0.010 & 0.016\\
\hline
Total               & 0.063 & 0.059 & 0.057 \\ \hline \hline
\end{tabular}
\end{table}



The systematic uncertainties are conservatively assumed to be fully correlated
except for those due to template statistics and the calibration of the 
individual analyses,
which are completely uncorrelated, and the MC
model systematic uncertainties which are partially correlated.
Assuming a fixed value of $0.7$ for \fzero, we find
\begin{equation}
\fplus=0.109\pm0.094\mathrm{ (stat) }\pm0.063\mathrm{ (syst) }
\end{equation}
using $\ell+$ jets events, and
\begin{equation}
\fplus=-0.089\pm0.154\mathrm{ (stat) }\pm0.059\mathrm{ (syst) }
\end{equation}
using dilepton events.  Combination of these results yields
\begin{equation}
\fplus=0.056\pm0.080\mathrm{ (stat) }\pm0.057\mathrm{ (syst) }.
\end{equation}
We also calculate a Bayesian confidence interval (using a flat prior
distribution which is non-zero only in the physically allowed 
region of $\fplus=0.0-0.3$) 
which yields
\begin{equation}
\fplus<0.23\mathrm{\ @\ }95\%\mathrm{\ \CL}
\end{equation}

 As seen in Fig.~\ref{fig_cost_ljets}(a), there is a deficit
of $\ell+$jets data events in the central region of $\cos\theta^*$.  We estimate the significance
of this effect by performing a likelihood ratio test to evaluate the 
goodness-of-fit for the best-fit model and find that the probability of 
obtaining a worse fit is 1.3\%.  We also evaluate the goodness of fit for 
the standard model hypothesis and find a fit probability of 0.8\% 
(statistical).  Thus we conclude that the discrepancy is not statistically 
significant.  We have studied the subset of our MC ensemble tests in 
which the mock data has a lower fit probability than the collider data does 
and find that our sensitivity to the value of \fplus in this subset
is the same as in the entire set of ensembles.

In summary, we have measured the fraction of right-handed $W$ bosons 
in \ttbar decays
in the \ljets and dilepton channels, and find 
$\fplus=0.056\pm0.080\mathrm{ (stat) }\pm0.057\mathrm{ (syst) }$.
This is the most precise 
measurement
of $\fplus$ to date and is consistent with the 
\SM prediction of $\fplus=3.6\times10^{-4}$~\cite{fischer1}.  

%
We thank the staffs at Fermilab and collaborating institutions, 
and acknowledge support from the 
DOE and NSF (USA);
CEA and CNRS/IN2P3 (France);
FASI, Rosatom and RFBR (Russia);
CAPES, CNPq, FAPERJ, FAPESP and FUNDUNESP (Brazil);
DAE and DST (India);
Colciencias (Colombia);
CONACyT (Mexico);
KRF and KOSEF (Korea);
CONICET and UBACyT (Argentina);
FOM (The Netherlands);
PPARC (United Kingdom);
MSMT (Czech Republic);
CRC Program, CFI, NSERC and WestGrid Project (Canada);
BMBF and DFG (Germany);
SFI (Ireland);
The Swedish Research Council (Sweden);
Research Corporation;
Alexander von Humboldt Foundation;
and the Marie Curie Program.


\begin{thebibliography}{99}
\bibitem{def_eta} Rapidity $y$ and pseudorapidity $\eta$ are defined
  as functions of the polar angle $\theta$ with respect to the proton
  beam and the parameter $\beta$ as $y(\theta,\beta)\equiv
  \frac{1}{2}\ln[(1+\beta\cos\theta)/(1-\beta\cos\theta)]$ and
  $\eta(\theta)\equiv y(\theta,1)$, where $\beta$ is the ratio of a
  particle's momentum to its energy.
\end{thebibliography}
\end{document}